\documentclass[%
reprint,
]{revtex4-2}

\usepackage{graphicx}
\usepackage{dcolumn}
\usepackage{bm}
\usepackage{tikz}
\usepackage{pgfplots}
\usepackage{subcaption}
\usepackage{hyperref}
\usepackage{float}
\usepackage{caption}
\usepackage{ragged2e}
\usepackage{amsmath}
\usepackage{hyperref}
\usepackage{multirow}
\usepackage{soul}

\begin{document}

\preprint{APS/123-QED}

\title{Characterization of dual-path coupled fluxonium qubits}

\title{Verifying the analogy between transversely coupled spin-1/2 systems and inductively-coupled fluxoniums}

\author{Wei-Ju Lin$^{1}$}
\altaffiliation{These authors contributed equally to this work}
\author{Hyunheung Cho$^{1}$}
\altaffiliation{These authors contributed equally to this work}
\author{Yinqi Chen$^{2}$}
\author{Maxim G. Vavilov$^{2}$}
\author{Chen Wang$^{3}$}
\author{Vladimir E. Manucharyan$^{1,4,}$}
\thanks{Author to whom any correspondence should be addressed. E-mail: vladimir.manucharyan@epfl.ch}

\affiliation{$^1$Department of Physics, University of Maryland, College Park, MD, USA}
\affiliation{$^2$Department of Physics, University of Wisconsin-Madison, Madison, WI, USA}
\affiliation{$^3$Department of Physics, University of Massachusetts-Amherst, Amherst, MA, USA}
\affiliation{$^4$Institute of Physics, Ecole Polytechnique Federale de Lausanne, Lausanne, Switzerland}

\date{\today}

\begin{abstract}
We report a detailed characterization of two inductively coupled superconducting fluxonium qubits for implementing high-fidelity cross-resonance gates. Our circuit stands out because it behaves very closely to the case of two transversely coupled spin-$1/2$ systems. In particular, the generally unwanted static $\rm{ZZ}$-term due to the non-computational transitions is nearly absent despite a strong qubit-qubit hybridization. Spectroscopy of the non-computational transitions reveals a spurious $LC$-mode arising from the combination of the coupling inductance and the capacitive links between the terminals of the two-qubit circuit. Such a mode has a minor effect on the present device, but it must be carefully considered for optimizing future multi-qubit designs.

\end{abstract}

\maketitle


\section{Introduction}
Engineering robust on-demand interactions between long-lived quantum bits (qubits) is important for enabling high-fidelity logical gates in future quantum computers and achievable by the outstanding tunability of superconducting circuits. In the case of superconducting qubits~\cite{Devoret2013, Wendin2017, Kjaergaard2020_review}, a common approach is to connect two frequency-detuned qubit circuits via a capacitor and activate the qubit-qubit interaction by microwave drives. 
A characteristic drawback of such a permanent connection is that it usually leads to a static $\rm{ZZ}$-term, which arises  from the generally uneven "pressure" on the four computational levels from the higher-energy non-computational levels. This interaction would induce coherent errors during single-qubit operations and lead to quantum cross-talk across the qubit register, which is one of the main challenges to build a superconducting quantum processor. This effect can be pronounced not only for weakly-anharmonic transmons \cite{wei2022hamiltonian, nguyen2024programmable, mitchell2021hardware, barends2019diabatic, kandala2021demonstration, negirneac2021high} but also for strongly-anharmonic fluxoniums \cite{xiong2022arbitrary, ficheux2021fast, dogan2023two, bao2022fluxonium}. The magnitude of the static $\rm{ZZ}$-term can be suppressed by a variety of tricks, from introducing more complex coupler elements \cite{mckay2016universal, yan2018tunable, mundada2019suppression, xu2020high, collodo2020implementation, kandala2021demonstration, sung2021realization, ding2023high, zhang2023tunable}, including fast-flux-tunable couplers \cite{bialczak2011fast}, to applying differential AC-stark shifts by off-resonantly driving the non-computational transitions \cite{xiong2022arbitrary}. However, both mitigation strategies increase the complexity of devices and control protocols.

The relatively large $\rm{ZZ}$-term in capacitively coupled fluxoniums is related to a general property of transition matrix elements of the charge operator. That is, they are proportional to the transition frequency \cite{nesterov2018microwave}. Consequently, even if the non-computational states are far detuned from the computational ones, their effect cannot be readily neglected. By contrast, the coupling of fluxoniums via a mutual inductance is governed by the flux operator, and the transition matrix elements do not generally grow with frequency \cite{zhang2023tunable, ma2024native}. Combining inductive and capacitive coupling can even lead to completely canceling the static $ZZ$ term \cite{nguyen2022blueprint}. We further explore the inductive coupling of fluxonium qubits for microwave-activated gates.

In this work we describe an inductively-coupled two-fluxonium device. The fluxonium circuits share a common junction, which acts as a linear mutual inductance. The drive is applied via the qubit's external antenna-like capacitors, which also couple both qubits to a 3D cavity for a joint dispersive readout. The antennas are convenient for creating wireless microwave drives, but they also create capacitive links between circuit terminals, and they need to be taken into account for accurate modeling of the interactions.

Our results can be formulated as follows. A small mutual inductance is indeed sufficient to hybridize qubits with relatively far detuned transitions ($f^{A}_{01} = 150~\rm{MHz}$, $f^{B}_{01} = 230~\rm{MHz}$) to create an on-demand cross-resonance (CR) interaction (${ZX}$-term) comparable to the Rabi rates of single-qubit gates
while the static $ZZ$-term is suppressed to a few kHz. The capacitance cross-talk does not affect the magnitude of the $ZX$-term but influences the static $ZZ$-term. In particular, it creates a previously overlooked 
$LC$ mode, which, in general, must be taken into account when identifying the spectrum and transition matrix elements of the coupled system. Observation and quantitative modeling of this mode, along with the demonstration of the nearly ideal transversely-coupled spin-$1/2$ Hamiltonian in the computational sub-space, constitute the main focus of the present work.

\section{Device and circuit model}
\begin{figure*}
    \centering
    \begin{tikzpicture}
        \node[inner sep=0] (image2) at (0,0) {\includegraphics[width=0.99\textwidth]{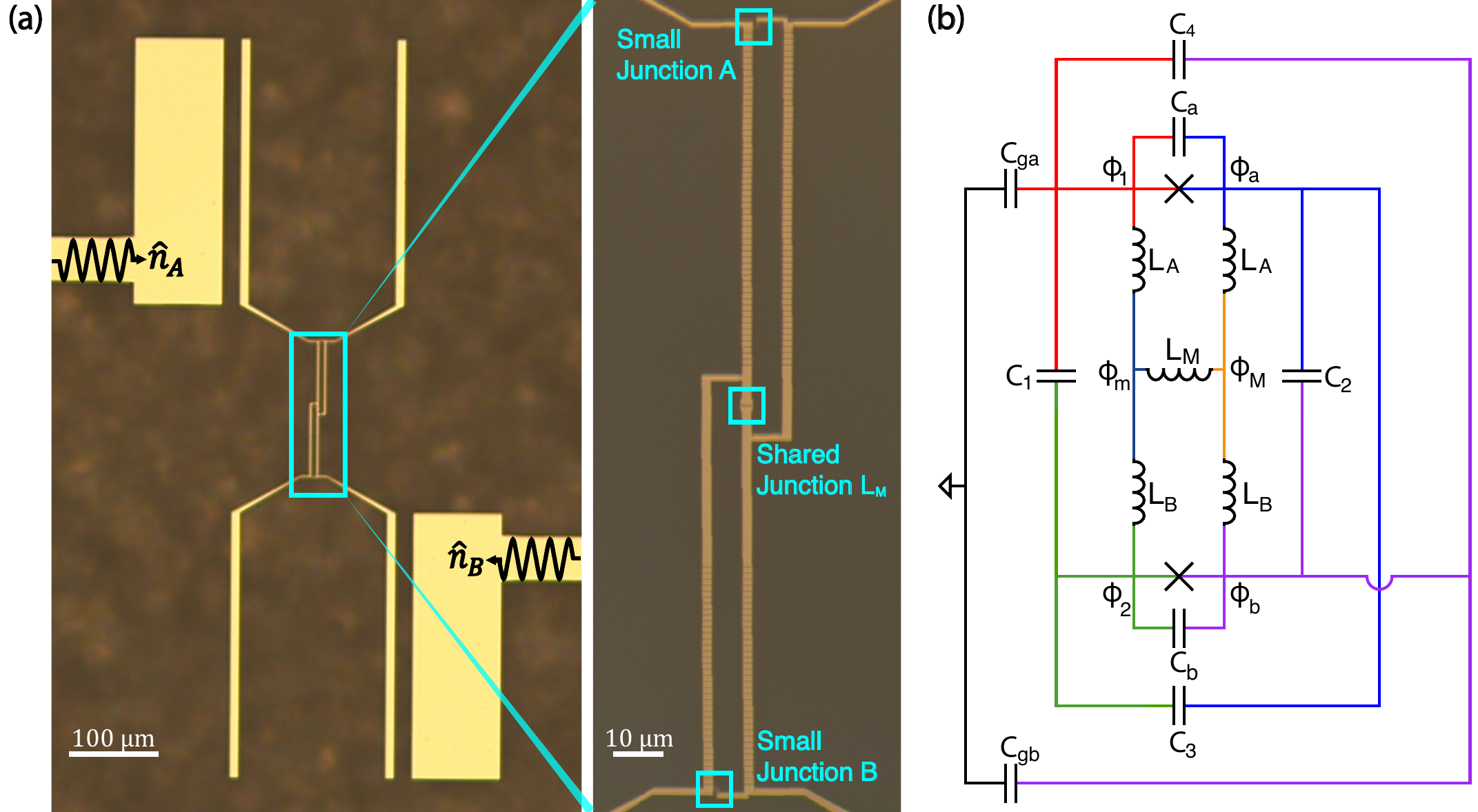}};
    \end{tikzpicture}
    \caption{\label{fig:opticalandcircuit} \justifying (a) The optical image of the two-fluxonium inductively coupled device. The joint readout cavity is not shown. The inset shows the mutual inductance $L_M$ and locations of the fluxoniums' small junctions. The left and right lead are designed to drive qubits A and B, respectively, illustrated by the microwave signal symbols with the charge operator $\hat{n}_{A(B)}$. (b) General circuit diagram of the device with false-color representation, taking into account both the inductive coupling and the capacitive cross-talk. Colors correspond to the node flux, $\phi$, while $L$ and $C$ denote inductors and capacitors, respectively.  }
\end{figure*}

\begin{figure*}
\includegraphics[width=0.99\textwidth]{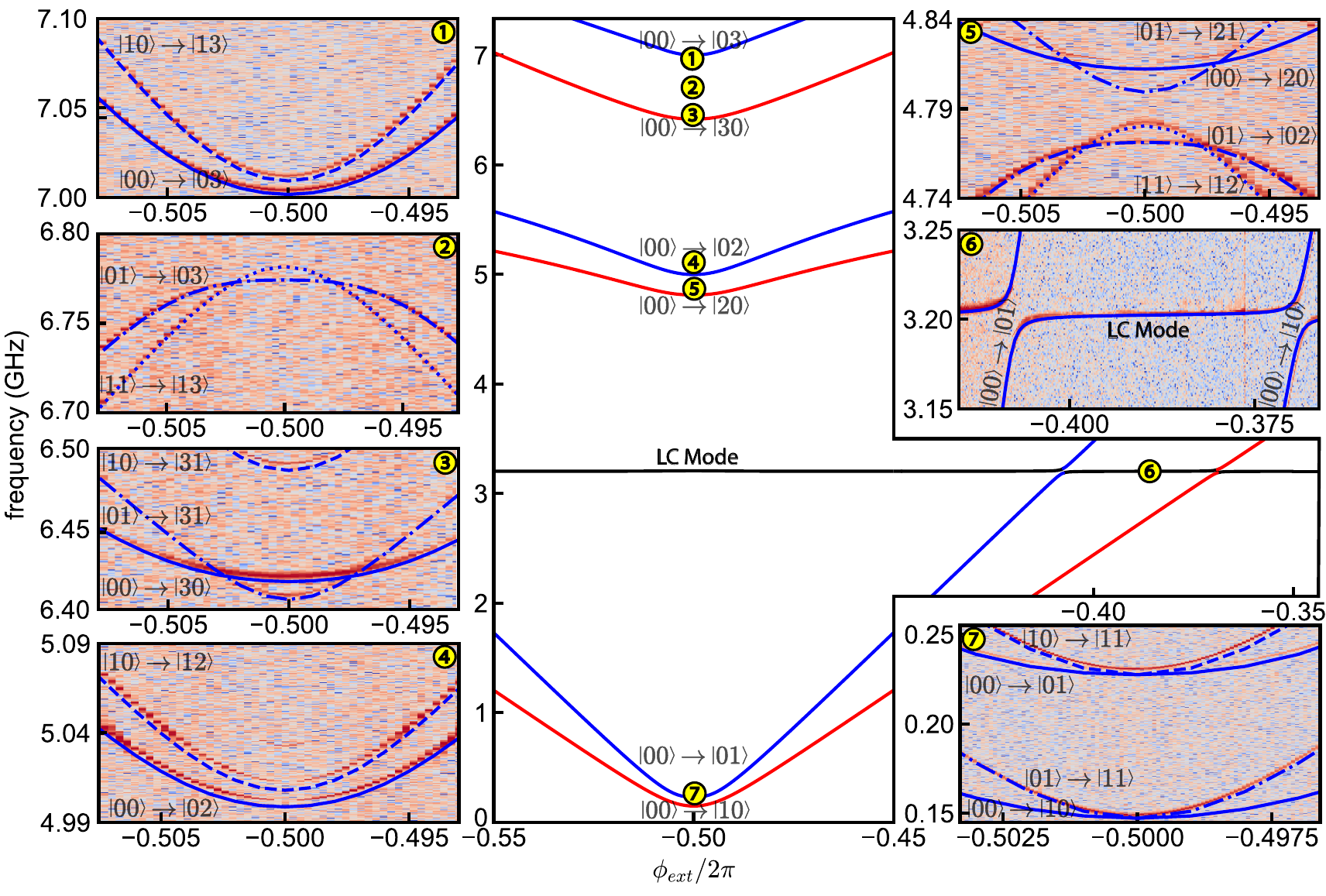}
\caption{\label{spectroscopy} \justifying Fit spectral lines for qubit A (red) and B (blue), assuming adiabatic indexing of the coupled system. The data is shown in insets indexed from \textcircled{1} to \textcircled{7}. The computational sub-space is shown in inset \textcircled{7}, the stray $LC$-mode and its interaction with the qubits is shown in inset \textcircled{6}, and the other insets show various non-computational transitions, all in good agreement with theory.
}
\end{figure*}
We start with constructing the full Hamiltonian and demonstrate the analogy between our system and transversely coupled spin-1/2 systems later with the property derived from the device parameters of the Hamiltonian. The Hamiltonian of two inductively and capacitively coupled fluxonium circuits can be written as follows\cite{nguyen2022blueprint}. 
\begin{equation}
\frac{\hat{\mathcal{H}}_{\text{fl}}}{h}
=\frac{\hat{\mathcal{H}}_A}{h}+\frac{\hat{\mathcal{H}}_B}{h}+J_C \hat{n}_A \hat{n}_B +J_L \hat{\varphi}_A \hat{\varphi}_B.
\end{equation}
The first two terms describe uncoupled qubits (labeled $\alpha = A, B$),
\begin{equation}
\frac{\hat{\mathcal{H}}_\alpha}{h}=4 E_{C, \alpha} \hat{n}_\alpha^2+\frac{1}{2} E_{L, \alpha} \hat{\varphi}_\alpha^2-E_{J, \alpha} \cos \left(\hat{\varphi}_\alpha-\phi_{\mathrm{ext}}\right),    
\end{equation}
where $E_J$, $E_C$, and $E_L$ are the charging, Josephson, and inductive energy, respectively. The capacitive and inductive coupling strengths are $J_C$ and $J_L$, and the charge operator is $2e\hat{n}$. The flux operator is $(\hbar/2e)\hat{\varphi}$ obey the canonical commutation relation $[\hat{n}, \hat{\varphi}] = i$; a global magnetic field (assuming equal area loops) creates a phase bias $\phi_{\rm{ext}}$. A more thorough circuit analysis for our device shown in Fig. \ref{fig:opticalandcircuit}(a) reveals the presence of an additional $LC$-mode as shown in Appendix \ref{sec:fullcircuitanalysis} based on Fig. \ref{fig:opticalandcircuit}(b), modifying the system Hamiltonian to
\begin{equation}
\begin{aligned}
  \frac{\hat{\mathcal{H}}_{\text{full}}}{h} = \frac{\hat{\mathcal{H}}_{\text{fl}}}{h} + f_{LC}\hat{a}^{\dagger}\hat{a} 
  -i\sum_{\alpha=A,B}g_{\alpha}\hat{n}_\alpha\left(\hat{a}-\hat{a}^\dagger \right). \label{totalhamiltonian1}
    \end{aligned}
\end{equation}
$\hat{a}$, and $\hat{a}^\dagger$ are the creation and annihilation operators of the bosonic mode. The Hamiltonian parameters for the device in question, obtained from the analysis of the measured spectrum, are summarized in Table \ref{tab:table1}. $f_{LC}$ represents the $LC$-mode frequency. Its coupling strength to the qubit $\alpha$ is $g_{\alpha}$. We also note that the sign of $J_L$ is always positive, while the sign of $J_C$ can either depend on the details of the capacitive cross-talk network. For simplicity of presentation, we are omitting the coupling to the readout mode at about $7.5~\rm{GHz}$, as it has a minor effect on the spectral properties of the two-qubit system.

\begin{table}[h]
\centering
\caption{\label{tab:table1}%
System parameters of the device}
\begin{ruledtabular}
\begin{tabular}{lcc}
\textrm{Parameter (GHz)} & \textrm{Qubit A} & \textrm{Qubit B} \\
\colrule
\quad \quad$E_{J,\alpha}$ & 5.59 & 6.27 \\
\quad \quad$E_{L,\alpha}$ & 0.76 & 1.16 \\
\quad \quad$E_{C,\alpha}$ & 0.98 & 0.99 \\
\quad \quad$g_{\alpha}$ & 0.18 & -0.21 \\
\quad \quad$f_{LC}$ & \multicolumn{2}{c}{3.22} \\
\quad \quad$J_C$ & \multicolumn{2}{c}{-0.038} \\
\quad \quad$J_L$ & \multicolumn{2}{c}{0.004} \\
\end{tabular}
\end{ruledtabular}
\end{table}

\section{System Characterization}

\noindent\textit{Fabrication.} 
Our device is fabricated on a 9 mm $\times$ 4 mm, and 430 $\mu$m thickness sapphire substrate. The resist is spin-coated onto the diced chip, starting with a layer of MMA resist followed by a layer of PMMA resist. An Elionix system is used for e-beam writing to pattern the design onto the resist-coated chip. After development, a mask is created to outline the qubit structures. The double-angle deposition is carried out using a Plassys deposition system, with aluminum deposited at two angles to form the qubit structures. The process concludes with a lift-off procedure to remove excess material, leaving the patterned qubit devices. More detailed fabrication process is described in \cite{somoroff2023millisecond}.\\

\noindent\textit{Setup.} The measurement setup is illustrated in \cite{lin202424days}. The chip is placed inside a 3D copper cavity with a resonance frequency of 7.475 GHz, and a linewidth $\kappa / 2\pi = 9$ MHz for dispersive joint readout. External driving is introduced into the cavity through two input ports, with the more strongly coupled output port used to measure the transmission signal. The cavity is thermally anchored in the base plate of a dilution refrigerator at a temperature of 14 mK. The device spectrum is measured using a standard two-tone experiment technique, where the cavity readout tone follows the system excitation tone. By fitting the spectrum at various external flux points using the numerical diagonalization of the Hamiltonian presented in Eq. \ref{totalhamiltonian1}, the qubit parameters are extracted, as shown in Table \ref{tab:table1}.\\


\noindent \textit{Spectroscopy.} Figure \ref{spectroscopy} shows the device spectrum with theoretical spectrum lines based on qubit parameters in Table \ref{tab:table1}, and the data is shown in the insets from \textcircled{1} to \textcircled{7}. Insets \textcircled{1} to \textcircled{3} show the transitions involving the $|3\rangle$ state, \textcircled{4} and \textcircled{5} illustrate the transitions involving the $|2\rangle$ state, \textcircled{6} highlights the $LC$-mode of the system, and inset \textcircled{7} focuses on the transitions involving the $|1\rangle$ state at half flux quantum. The black line in the main figure represents the spurious $LC$-mode of the system. In the main figure, each index marks the corresponding region for the insets. The two-tone spectroscopy data in Fig. \ref{spectroscopy} reveals that the $|0\rangle-|1\rangle$ transition frequencies for qubits A and B are 0.15 GHz and 0.23 GHz, respectively. The $|1\rangle-|2\rangle$ transition frequencies are near 4.66 GHz and 4.78 GHz, giving the large anharmonicities of 4.51 GHz for qubit A and 4.54 GHz for qubit B. The relatively large anharmonicity compared with the qubit frequency indicates the leakage to higher energy levels while driving qubits can be significantly suppressed.\\

\begin{figure}
\includegraphics[width=0.49\textwidth]{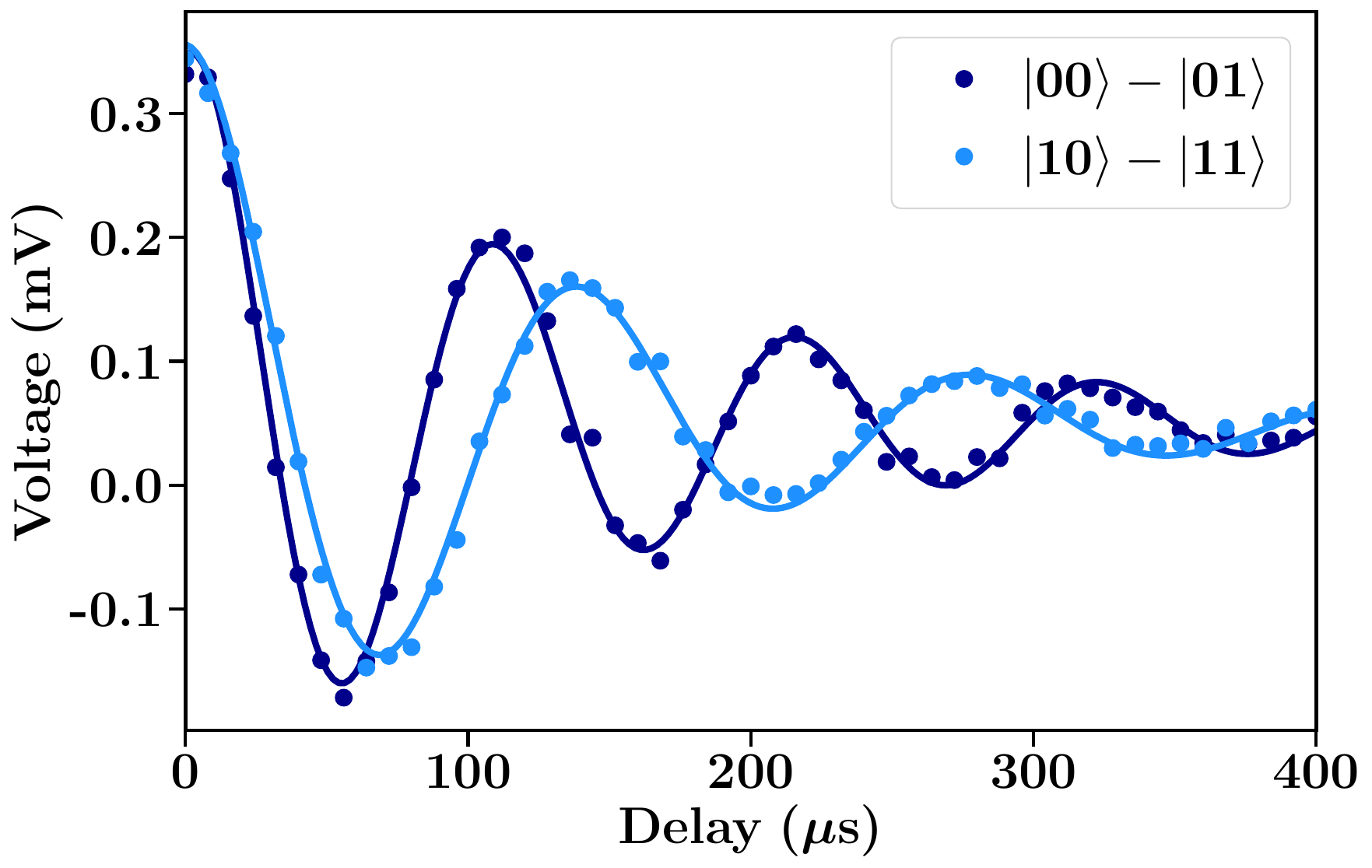}
\caption{\label{staticZZ}  
\justifying Ramsey fringe of qubit B conditioned on the state of qubit A. The experiment involves measuring the static $ZZ$ coupling strength with qubit $A$ in the $|0\rangle$ state (dark blue) and in the $|1\rangle$ state (blue). The difference of the two Ramsey frequencies gives the static $ZZ$-shift, $\xi^{\rm{static}}_{ZZ} = 2~\rm{kHz}$. The measurement relies on a relatively long decoherence time of the two-qubit system.
}
\end{figure}

\begin{figure*}
    \centering
    \begin{tikzpicture}
        \node[inner sep=0] (image1) at (0, 0) {\includegraphics[width=0.99\textwidth]{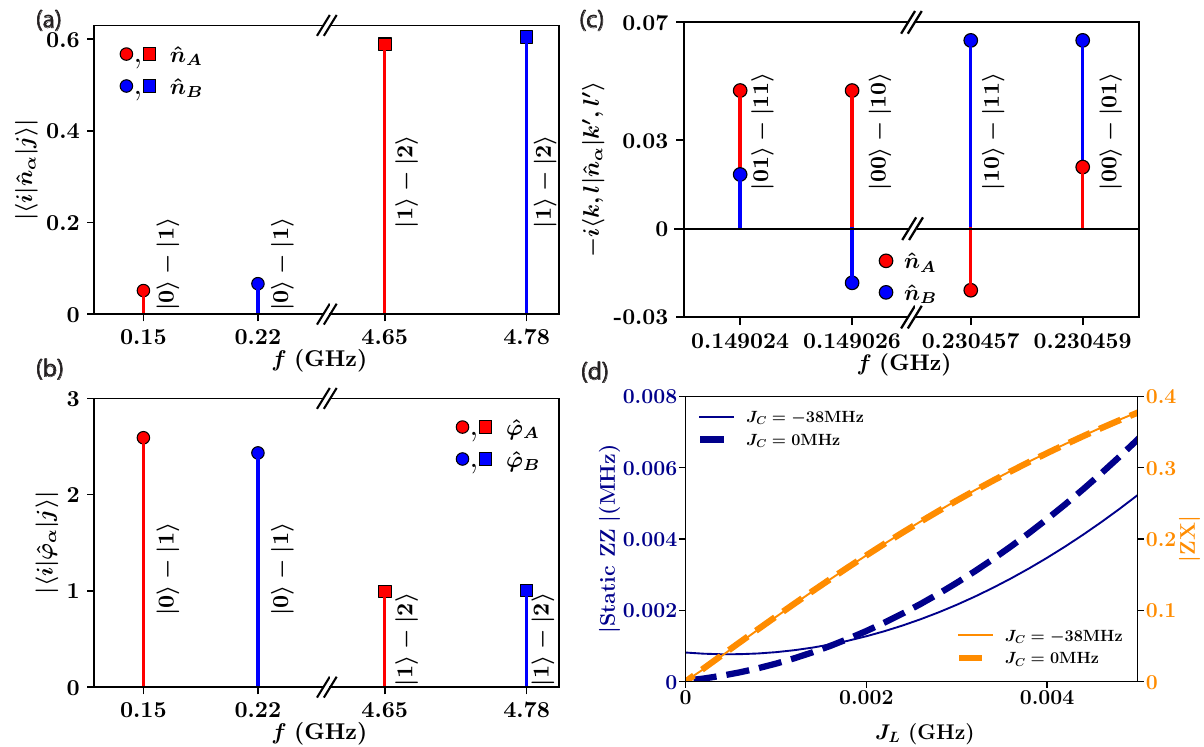}};
    \end{tikzpicture}
    \caption{\label{fig:matrixelementsall} \justifying The charge (a) and phase (b) matrix elements of the two inidividual qubits without coupling. The qubit parameters are listed in Table I. (c) Two-qubit charge matrix elements for computational transitions. Note, their almost symmetric structure is nearly identical to those of a system of two transversly-coupled spin-half systems. The qubit frequencies are extracted from Ramsey measurements. (d) Numerical values of the static $ZZ$ and the $ZX$-term. Note the coupling capacitances have no effect on the $ZX$-term and a minor effect on the $ZZ$-term.}
    
\end{figure*}

\noindent \textit{Static ${ZZ}$-term.} Figure \ref{fig:matrixelementsall}(a) and (b) present the numerical charge and phase matrix elements, respectively, for the $|0\rangle - |1\rangle$ and $|1\rangle - |2\rangle$ transitions using our bare qubit parameters without any qubit-qubit coupling. We selected the $|1\rangle - |2\rangle$ transition to represent the non-computational space because it exhibits much larger values than other non-computational transitions. Notably, the ratio $\left|\langle 0 | \hat{n}_{\alpha} | 1 \rangle / \langle 1 | \hat{n}_{\alpha} | 2 \rangle \right|< 1$ implies that achieving strong hybridization of the $|01\rangle$ and $|10\rangle$ states without involving higher energy states is challenging, potentially leading to $ZZ$ interactions if higher energy states are involved. Conversely, the ratio $\left|\langle 0 | \hat{\varphi}_{\alpha} | 1 \rangle / \langle 1 | \hat{\varphi}_{\alpha} | 2 \rangle \right| > 1$ suggests that the influence of higher energy states is suppressed. Combining with the large frequency detuning between the qubit A (B) transition and the $|1\rangle - |2\rangle$ transition of qubit B (A), this enables strong hybridization between the $|01\rangle$ and $|10\rangle$ states with negligible $ZZ$ interaction, resembling a pair of transversely coupled spin-1/2. 
To be specific, we next analyze the scheme using our device parameters. In spin-based qubit representations, interactions are typically expressed in terms of Pauli operators, such as \( J \sigma_x \sigma_x \). In contrast, fluxonium qubits are described using charge (\( \hat{n} \)) and phase (\( \hat{\varphi} \)) operators, leading to capacitive coupling terms \( J_C \hat{n}_A \hat{n}_B \) and inductive interactions \( J_L \hat{\varphi}_A \hat{\varphi}_B \). To determine the coupling strength in the spin representation, the charge and phase matrix elements from Figure \ref{fig:matrixelementsall}(a), (b) must be considered. The capacitive and inductive coupling strengths can be evaluated using the following expressions: \( |J_C \langle 0 | \hat{n}_A | 1 \rangle \langle 0 | \hat{n}_B | 1 \rangle| \) and \( |J_L \langle 0 | \hat{\varphi}_A | 1 \rangle \langle 0 | \hat{\varphi}_B | 1 \rangle |\). With \( J_C = -38 \, \text{MHz} \) and \( J_L = 4 \, \text{MHz} \), the capacitive coupling contributes approximately \( 0.1 \, \text{MHz} \), while the inductive coupling dominates at around \( 20 \, \text{MHz} \) due to the large phase matrix elements, as shown in Figure \ref{fig:matrixelementsall}(b). The coupling to higher-level transitions due to the large \( |\langle 1 | \hat{n}_A | 2 \rangle| \) and \( |\langle 1 | \hat{n}_B | 2 \rangle |\) in Figure \ref{fig:matrixelementsall}(a) is significantly suppressed considering \( |J_C \langle 1 | \hat{n}_A | 2 \rangle \langle 0 | \hat{n}_B | 1 \rangle |\) and \( |J_C \langle 0 | \hat{n}_A | 1 \rangle \langle 1 | \hat{n}_B | 2 \rangle| \approx 1 \, \text{MHz} \), which is still much smaller than the \( 4.5 \, \text{GHz} \) anharmonicity.

As a result, our system utilizing direct coupling achieves a low static $ZZ$ value, approximately 2 kHz. To accurately extract this value, we conduct Ramsey experiments on qubit $B$ while setting qubit $A$ to either its ground state or its excited state as shown in Fig. \ref{staticZZ}. The difference in the frequencies between the two Ramsey fringes is given by $\zeta_{ZZ}^{\rm{static}}$. We then fit the oscillations to extract the two transition frequencies with a decaying sinusoidal function, and subtract the two frequencies to get the value of $\zeta_{ZZ}^{\rm{static}}$:
\begin{equation}
\xi_{ZZ}^{\rm{static}} = f_{|00\rangle-|01\rangle}-f_{|10\rangle-|11\rangle}.
\end{equation}

\section{Truncating to the 4 computational states}
\subsection{Cross-resonance interaction}

Having established the accurate circuit model via spectroscopy, we proceed with truncating the system to its four lowest computational energy states:
\begin{equation}
\frac{\mathcal{\hat{H}}_{\mathrm{eff}}}{h} = 
    \frac{f_{01}^A}{2} \hat{\sigma}_z^A + \frac{f_{01}^B}{2} \hat{\sigma}_z^B + \frac{\xi_{ZZ}^{\rm{static}}}{4} \hat{\sigma}_z^A \hat{\sigma}_z^B +\frac{\mathcal{\hat{H}}_{\mathrm{drive}}}{h},
    \label{eq:totalHamiltonian}
\end{equation}
where $h$ is the Plank constant. $f_{01}^{A,B}$ is the frequency of qubits $A$ and $B$, and $\hat{\sigma}_{z}^{A,B}$ are the corresponding Pauli operators. The quantity $\xi_{ZZ}^{\text{static}}$ represents the static $ZZ$ interaction rate measured in Fig. \ref{staticZZ}. The microwave drive Hamiltonian in the computational subspace reduces to\begin{equation}\label{eq:DriveEffHamiltonian}
        \frac{\mathcal{\hat{H}}_{\text {drive}}}{h} = \beta(t)
        \left( \xi_{A}^{+} \hat{\sigma}_x^A + \xi_{A}^{-} \hat{\sigma}_x^A \hat{\sigma}_z^B
        + \xi_{B}^{+} \hat{\sigma}_x^B + \xi_{B}^{-} \hat{\sigma}_z^A \hat{\sigma}_x^B \right),
\end{equation}
where $\beta(t)$ describes the time-dependence of the drive oscillating with frequency $f_d$ equal to one of the qubit frequencies and a smooth amplitude modulation.
The real factors $\xi_{A}^{+}$, $\xi_{A}^{-}$, $\xi_{B}^{+}$, and $\xi_{B}^{-}$ represent effective $XI$, $XZ$, $IX$, and $ZX$ interactions, respectively, which can be written in terms of purely imaginary charge matrix elements as
\begin{equation}
    \begin{split}
    i \xi_{A}^{\pm} & = \langle 00 | \epsilon_{A} \hat{n}_{A} + \epsilon_{B} \hat{n}_{B} | 10 \rangle \pm \langle 01 | \epsilon_{A} \hat{n}_{A} + \epsilon_{B} \hat{n}_{B} | 11 \rangle,\\
    i \xi_{B}^{\pm} & =  \langle 00 | \epsilon_{A} \hat{n}_{A} + \epsilon_{B} \hat{n}_{B} | 01 \rangle \pm \langle 10 | \epsilon_{A} \hat{n}_{A} + \epsilon_{B} \hat{n}_{B} | 11 \rangle.
    \label{eq:xi_pm}
    \end{split}
\end{equation}
The drive amplitudes $\epsilon_{A}$ and $\epsilon_{B}$ are real and represent two in-phase drives on the two fluxoniums.

\subsection{Two-qubit charge matrix elements}

\begin{table}
  \caption{\label{tab:MatrixElements} \justifying Charge matrix elements of the transitions inside computational space for charge drives on the two fluxoniums.}
  \begin{tabular}{lcc}
  \hline \hline \multirow{1}{*}{$-i\langle kl|\hat{n}_{j}| k'l'\rangle$} & $\hat{n}_{A}$ & $\hat{n}_{B}$ \\
  \hline
  $\quad| 00\rangle-| 10\rangle$ & 0.046865 & -0.018371 \\
  $\quad| 01\rangle-| 11\rangle$ & 0.046871 & 0.018397 \\  
  $\quad| 00\rangle-| 01\rangle$ & 0.020908 & 0.063885 \\
  $\quad| 10\rangle-| 11\rangle$ & -0.020883 & 0.063873 \\
  \hline \hline
  \end{tabular}
\end{table}

In this section, we describe the charge matrix elements of the transitions within the computational space as shown in 
Fig.~\ref{fig:matrixelementsall}(c) and Table. \ref{tab:MatrixElements}.  These matrix elements define the coefficients~\ref{eq:xi_pm} of the time-dependent Hamiltonian~\eqref{eq:DriveEffHamiltonian}. 
Specifically, we can treat these coefficients as four control knobs: (i) the local drive $\epsilon_A$ at the frequency $f_{01}^A$, (ii) the local drive $\epsilon_B$ at the frequency $f_{01}^A$, (iii) the local drive $\epsilon_B$ at the frequency $f_{01}^B$, and (iv) the local drive $\epsilon_A$ at the frequency $f_{01}^B$. These control knobs provide tuning the $XI$, $XZ$, $IX$, and $ZX$ interaction rates individually, simplifying the operational complexity of cross-resonance gates.
 
A direct drive $\epsilon_A$ ($\epsilon_B$) at the frequency $f_{01}^A$ ($f_{01}^B$) causes the Rabi oscillations in qubit A (B) subspace with the Rabi frequency determined by
the matrix elements $\langle 01 | \hat{n}_{A} | 11 \rangle$ and $\langle 00 | \hat{n}_{A} | 10 \rangle$ ($ \langle 10 | \hat{n}_{B} | 11 \rangle$ and $\langle 00 | \hat{n}_{B} | 01 \rangle$).  These pairs of matrix elements have the same magnitude and sign regardless of the state of the other qubit, the condition necessary for single-qubit gates.  
On the other hand, a cross-resonant drive $\epsilon_A$ ($\epsilon_B$) at the frequency $f_{01}^B$ ($f_{01}^A$) generates rotation of qubit B (A) state with the Rabi frequency proportional to the matrix elements 
$\langle 10 | \hat{n}_{A} | 11 \rangle$ and $\langle 00 | \hat{n}_{A} | 01 \rangle$ ($\langle 01 | \hat{n}_{B} | 11 \rangle$ and $\langle 00 | \hat{n}_{B} | 10 \rangle$).  Each pair of  these matrix elements has the same magnitudes but opposite signs, 
resulting in state rotations in opposite directions, dependent on the state of qubit A (B). This behavior is suitable for conditional two-qubit gates. 
The four matrix elements relevant to the cross-resonant drive have magnitudes comparable to those relevant to the direct drive, indicating strong hybridization of computational states. The similarity of matrix elements under strong hybridization indicates the feature of a transversely coupled spin-1/2 system, providing a simple scenario for gate operations with low complexity. For reference, we have demonstrated high-fidelity cross-resonance gates accordingly on this device \cite{lin202424days}.

\subsection{Magnitude of the $\rm{ZX}$-term}

We further investigate the resulting magnitude of the $ZX$-interaction under the same drive amplitude for single-qubit gates. We evaluate this entangling strength using the following expression:
\begin{equation}
\begin{split}
|ZX|= & \left| \frac{\xi_B^{-}\left(\epsilon_A=\epsilon, \epsilon_B=0\right)}{\xi_B^{+}\left(\epsilon_A=0, \epsilon_B=\epsilon\right)} \right| \\
= &
\left| \frac{\langle 00 | \hat{n}_{A} | 01 \rangle - \langle 10 | \hat{n}_{A} | 11 \rangle}{\langle 00 | \hat{n}_{B} | 01 \rangle+\langle 10 | \hat{n}_{B} | 11 \rangle} \right|,
\end{split}
\end{equation}
The numerator represents the cross-resonant drive with $\epsilon_A$ turned on, and $\epsilon_B$ turned off, and the denominator represents the direct drive with $\epsilon_A$ turned off, and $\epsilon_B$ turned on. Figure \ref{fig:matrixelementsall}(d) shows $|ZX|$ as a function of the inductive coupling strength $J_L$ along with the static $ZZ$ phase accumulation rate.
We observe that the $ZX$ term grows linearly while the $ZZ$ term grows quadratically in $J_L$. This facilitates a significant separation of the two frequency scales. For the device in question, a purely inductive coupling of $J_L = 4~\rm{MHz}$ (neglecting the capacitive effects, that is setting $J_C =0$ and $g_{A(B)} =0$) results in a relatively small $\xi_{ZZ}^\mathrm{static} = 5~\rm{kHz}$, while the $ZX$ term can readily be close to 0.4. We also observe that the capacitive cross-talk does not affect the $ZX$-term at all for the device in question and helps suppress the $ZZ$-term a bit. The measured value of $2~\rm{kHz}$ qualitatively agrees with the theoretical prediction. 
For comparison, previous work \cite{dogan2023two} utilizing capacitive coupling also employs direct coupling schemes designed for cross-resonance gates in charge-driven fluxonium systems. A key difference is the resulting static ZZ interaction, where the capacitive coupling case leads to a static ZZ of 900 kHz, while our inductively coupled system achieves a significantly lower value of 2 kHz.

\section{Summary}

We have implemented a system of two inductively coupled but capacitively driven fluxonium qubits. The device behaves as a nearly ideal transversely-coupled spin-1/2 system and is thus well suited for all-microwave fixed-frequency cross-resonance two-qubit gates~\cite{Paraoanu2006, Rigetti2010}. The CR gate is performed by microwave excitation applied to the control qubit at the frequency of the target qubit and was successfully implemented first in flux qubits~\cite{deGroot2010,Chow2011}, and later in   transmons~\cite{Chow2012, Corcoles2013, Takita2016, Sheldon2016b, Jurcevic2021, kandala2021demonstration}. 

It is interesting to note that the values of the qubit frequencies $f_{01}^{A}$ and $f_{01}^{B}$ are relatively far from each other than in transmon experiments, which provides more freedom in the choice of circuit parameters. The strength of $ZX$-interaction is proportional to $J_L/|f_{01}^{A} - f_{01}^{B}|$ just like in the case of purely two-level systems. The static $ZZ$-term is suppressed into the low-kHz range thanks to the combination of a large frequency detuning of the non-computational states and the property of the flux matrix elements in Fig.\ref{fig:matrixelementsall}(b). Tuning the coupling capacitance can help suppress the $ZZ$ -term to zero, but it also needs to consider the corresponding change of $g_{A}$, $g_{B}$, and $f_{LC}$ of the stray $LC$-mode. This case would be a liability in terms of coherent and incoherent errors during gate operations. Although $LC$ mode does not introduce significant detrimental effects in our current design, additional spurious $LC$ modes could introduce unwanted interactions in multi-qubit architectures. These can be mitigated by ensuring sufficient detuning from qubit frequencies and small enough coupling constants to qubits, which can be achieved by minimizing the capacitive coupling between qubits. Since inductive coupling is preferred for qubit-qubit interactions, minimizing capacitive links aligns with the goal of reducing the unwanted coupling strength $J_C$. Another strategy to eliminate $LC$ modes entirely is the implementation of geometric mutual inductance, which could be achieved by the assistance of air bridges or flip-chip designs.

\section{Data Availability Statemenet}
The data that support the findings of this study are available upon reasonable request from the authors.

\begin{acknowledgments}
This research was supported by the ARO HiPS (contract No. W911-NF18-1-0146) and GASP (contract No. W911-NF23-10093) programs.
\end{acknowledgments}
\appendix

\section{Full circuit model analysis}\label{sec:fullcircuitanalysis}
To achieve the Hamiltonian in Eq. \ref{totalhamiltonian1}, we follow the node-flux analysis \cite{vool2017introduction}. We begin by establishing the Lagrangian for the capacitive, Josephson junction, and linear inductive components. This is followed by applying the Legendre transformation to derive the Hamiltonian.

\begin{equation}
    \mathcal{L}_{\text{tot}} = \mathcal{L}_{\text{cap}} + \mathcal{L}_{\text{ind}} + \mathcal{L}_{\text{JJ}}
\end{equation}

First, we set up the Lagrangian for the capacitance part:

\begin{equation}
\begin{aligned}
    \mathcal{L}_{\text{cap}} = \frac{1}{2} \left[ 
    \vphantom{\dot{\phi}_b^2} 
    \right. & C_a(\dot{\phi}_a - \dot{\phi}_1)^2 + C_b(\dot{\phi}_b - \dot{\phi}_2)^2 \\
     + & C_1(\dot{\phi}_1 - \dot{\phi}_2)^2 + C_2(\dot{\phi}_a - \dot{\phi}_b)^2 \\
     + & C_3(\dot{\phi}_a - \dot{\phi}_2)^2 + C_4(\dot{\phi}_b - \dot{\phi}_1)^2 \\
     + & C_{ga} \dot{\phi}_1^2 + C_{gb} \dot{\phi}_b^2 \left. \vphantom{\dot{\phi}_b^2} \right]
\end{aligned}
\end{equation}

Each variable $\phi_1, \phi_2, \phi_a, \phi_{b,} \phi_{m,} \text { and } \phi_M$ corresponds to specific node fluxes in the circuit, and their the time derivative of each term, as depicted in Fig. \ref{fig:opticalandcircuit} (b). 

Then, we define the new variables:
\begin{align}
    &\phi_A = \phi_a - \phi_1, \\
    &\phi_B = \phi_2 - \phi_b, \\
    &\phi_{LC} = \phi_a + \phi_1 - \phi_b - \phi_2, \\
    &\phi_{S} = \phi_a + \phi_1 + \phi_b + \phi_2,
\end{align}

and their time-derivatives:
\begin{align}
    &\dot{\phi}_A = \dot{\phi}_a - \dot{\phi}_1, \\ 
    &\dot{\phi}_B = \dot{\phi}_2 - \dot{\phi}_b, \\
    &\dot{\phi}_{LC} = \dot{\phi}_a + \dot{\phi}_1 - \dot{\phi}_2 - \dot{\phi}_b, \\
    &\dot{\phi}_{S} = \dot{\phi}_a + \dot{\phi}_1 + \dot{\phi}_2 + \dot{\phi}_b.
\end{align}

Therefore, the new Lagrangian for the capacitance part can be written as follows.
\begin{equation}
\begin{aligned}
\mathcal{L}_{\text {cap }}&=\frac{1}{2}  \left(C_a \dot{\phi}_A^2+C_b \dot{\phi}_B^2\right) \\
& +\frac{1}{8} C_1\left(\dot{\phi}_A+\dot{\phi}_B-\dot{\phi}_{L C}\right)^2 \\
& +\frac{1}{8} C_2\left(\dot{\phi}_A+\dot{\phi}_B+\dot{\phi}_{L C}\right)^2 \\
& +\frac{1}{8} C_3\left(\dot{\phi}_A-\dot{\phi}_B+\dot{\phi}_{L C}\right)^2 \\
& +\frac{1}{8} C_4\left(-\dot{\phi}_A+\dot{\phi}_B+\dot{\phi}_{L C}\right)^2 \\
& +\frac{1}{32} C_{g b}\left(2 \dot{\phi}_B+\dot{\phi}_{L C}-\dot{\phi}_S\right)^2 \\
& +\frac{1}{32} C_{g a}\left(-2 \dot{\phi}_A+\dot{\phi}_{L C}+\dot{\phi}_S\right)^2
\end{aligned}
\end{equation}

Second, we need to set up the Lagrangian for the Josephson junctions, which can be written as:
\begin{equation}
\begin{aligned}
    \mathcal{L}_{\text{JJ}} &= E_{J A} \cos (\frac{2 \pi}{\Phi_o}\left(\phi_A+\Phi_{\text{ext}, A}\right)) \\
    &+E_{J B} \cos (\frac{2 \pi}{\Phi_o}\left(\phi_B+\Phi_{\text{ext}, B}\right))
\end{aligned}
\end{equation}

Finally, we set up the Lagrangian for the linear inductance part:
\begin{equation}
    \begin{aligned}
        \mathcal{L}_{\text{ind}} = &- \frac{(\phi_1 - \phi_m)^2 + (\phi_a - \phi_M)^2}{2L_A}\\
        &- \frac{(\phi_2 - \phi_m)^2 + (\phi_b - \phi_M)^2}{2L_B} \\
        &-\frac{(\phi_m - \phi_M)^2}{2L_M},
    \end{aligned}
\end{equation}

We can eliminate $\phi_m$ and $\phi_M$ from the equations of motion:
\begin{equation}
    \frac{d}{d t} \frac{\partial \mathcal{L}_{\text{tot}}}{\partial \dot{\phi}_{k}}=\frac{\partial \mathcal{L}_{\text{tot}}}{\partial \phi_{k}}=0,
\end{equation}
where $k = m$, or $M$. 

Using these relations for $\phi_m$ and $\phi_M$, we can rewrite the Lagrangian as follows:
\begin{equation}
\begin{aligned}
 \mathcal{L}_{\text{ind}}= &-\frac{(2 L_B+L_M) \phi_A^2}{8 L_A L_B+4 L_A L_M+4 L_B L_M}\\
 &-\frac{L_M \phi_A \phi_B}{4 L_A L_B+2 L_A L_M+2 L_B L_M} \\
 &- \frac{(2 L_A+L_M) \phi_B^2}{8 L_A L_B+4 L_A L_M+4 L_B L_M}\\
 &-\frac{\phi^2_{LC}}{4(L_A+L_B)}.
\end{aligned}
\end{equation}

Applying the approximation,
\begin{equation}
    L_A, L_B \gg L_M,
\end{equation}
we obtain the following Lagrangian for the linear inductance part:
\begin{equation}
\begin{aligned}
\mathcal{L}_{\text{ind}}=&-\frac{\phi^2_A}{2\left(2 L_A\right)}-\frac{\phi^2_B}{2\left(2 L_B\right)} - \frac{\phi^2_{L C}}{2\left(2\left(L_A+L_B\right)\right)}\\
&-\frac{L_M \phi_A \phi_B}{\left(2 L_A\right)\left(2 L_B\right)}.    
\end{aligned}
\end{equation}

Therefore, we get the total Lagrangian,
\begin{equation}
    \mathcal{L}_{\text{tot}} = \mathcal{L}_{\text{cap}} + \mathcal{L}_{\text{ind}} + \mathcal{L}_{\text{JJ}}.
\end{equation}

The charge variables can be found following the relation $q_n \equiv \partial \mathcal{L} / \partial \dot{\phi}_n$. The Hamiltonian can be found via the Legendre transformation, giving:
\begin{equation}
    \mathcal{H} = \sum_{i=A,B,LC}q_i\dot{\phi}_i - \mathcal{L}_{\text{tot}}
\end{equation}

Therefore, we obtain the Hamiltonian:
\begin{equation}
    \begin{aligned}
        \mathcal{H} &= \frac{q_A^2}{2 C_A}+\frac{q^2_B}{2 C_B}+\frac{q^2_{L C}}{2 C_{L C}}+\frac{J_C q_A q_B}{4e^2}\\
        &+ \frac{\phi^2_A}{2\left(2 L_A\right)}+\frac{\phi^2_B}{2\left(2 L_B\right)} + \frac{\phi^2_{L C}}{2\left(2\left(L_A+L_B\right)\right)}+\frac{4e^2 J_L \phi_A \phi_B}{\hbar^2} \\
        &+\frac{{g_{A} q_A q_{L C}}}{4e^2}+\frac{{g_{B} q_B q_{L C}}}{4e^2} \\ 
        &- E_{J A} \cos \left(\phi_A-\phi_{e x t}\right)-E_{J B} \cos \left(\phi_B-\phi_{e x t}\right).
    \end{aligned}
\end{equation}

There are additional terms of $q_S$, which is a free variable without $\phi_S$. 
In the final step, the variables are promoted to quantum operators:
\begin{equation}
    \begin{aligned}
        q & \rightarrow \hat{q}=2 e \hat{n} \\
        \Phi & \rightarrow \hat{\Phi}=\frac{\hbar}{2 e} \hat{\varphi}
    \end{aligned}
\end{equation}

This yields the final Hamiltonian in Eq. \ref{totalhamiltonian1}, agreeing well with the results from scQubits package \cite{groszkowski2021scqubits}. The detailed forms of the coupling and capacitance terms are provided below.

The inductive coupling constant between qubit A and B, $J_L$, is as follows:
\begin{equation}
    J_L=\left( \frac{\hbar}{2e}\right)^2 \frac{L_M}{\left(2 L_A\right)\left(2 L_B\right)}
\end{equation}

The capacitive coupling constant between qubit A and B, $J_C$, is as follows:
\begin{equation}
\begin{aligned}
    J_C &=(4e^2) \left((C_3 C_{ga}C_{gb} - (Cga + Cgb)(C_1 C_2 - C_3 C_4
      )\right)/\\
& ((C_3 C_4 C_a + 
      C_2 C_3 (C_4 + C_a) + (C_2 + C_3) C_4 C_b \\
      &+ (C_2 + C_3 + C_4) C_a C_b + 
      C_1 (C_3 + C_a) (C_4 + C_b) \\
      &+C_1 C_2 (C_3 + C_4 + C_a + C_b)) C_{ga} \\
      &+ (C_3 C_4 C_a + C_3 C_4 C_b + 
      C_3 C_a C_b + C_4 C_a C_b \\
      &+ C_3 C_a C_{ga} + C_3 C_b C_{ga} + C_a C_b C_{ga}\\
      &+ 
      C_2 (C_3 + C_b) (C_4 + C_a + C_{ga}) \\
      &+ C_1 (C_3 + C_a) (C_4 + C_b + C_{ga}) \\
      &+ C_1 C_2 (C_3 + C_4 + C_a + C_b + C_{ga})) C_{gb})
\end{aligned}
\end{equation}
The coupling constants  $J_L$ and  $J_C$ reflect the design choices of our circuit. The inductive coupling  $J_L$  is determined by the ratio between the shared inductance  LM  and the linear inductances of qubits A and B,  $L_A$ and  $L_B$.
In contrast, the capacitive coupling  $J_C$  arises from the overall circuit geometry and contributions from all capacitive links. To suppress  $J_C$ , we design the circuit layout and the device geometry according to  the simulations  using electromagnetic software.

The capacitive coupling constant between qubit A and $LC$-Mode, $g_{A}$, is as follows:
\begin{equation}
\begin{aligned}
g_{A}&=(4e^2)(C_1 C_4 C_{ga} + (C_1 - C_3 + C_4) C_b C_{ga}\\
&+ C_b (-C_3 + C_4 + C_{ga}) C_{gb} + 
   C_1 (C_4 + C_b + C_{ga}) C_{gb} \\
   &- 
   C_2 (C_3 + C_b) (C_{ga} + C_{gb}))/\\
   &((C_3 C_4 C_a + C_4 C_a C_b + C_3 (C_4 + C_a) C_b \\
   &+
       C_2 (C_4 + C_a) (C_3 + C_b)) C_{ga} \\
       &+ 
   C_1 ((C_3 + C_a) (C_4 + C_b) \\
   &+ C_2 (C_3 + C_4 + C_a + C_b)) C_{ga} \\
   &+ 
   C_1 ((C_3 + C_a) (C_4 + C_b + C_{ga}) \\
   &+ 
      C_2 (C_3 + C_4 + C_a + C_b + C_{ga})) C_{gb} \\
      &+ (C_a C_b (C_4 + C_{ga}) + 
      C_2 (C_3 + C_b) (C_4 + C_a + C_{ga})\\
      &+ 
      C_3 (C_a C_b + C_4 (C_a + C_b) + (C_a + C_b) C_{ga})) C_{gb}) 
\end{aligned}    
\end{equation}
The capacitive coupling constant between qubit B and $LC$-Mode, $g_{B}$, is as follows:
\begin{equation}
\begin{aligned}
g_{B} &=(4e^2) (-((C_2 C_4 + (C_2 - C_3 + C_4) C_a) C_{ga}) \\
&- (C_a (-C_3 + C_4 + C_{ga}) + 
      C_2 (C_4 + C_a + C_{ga})) C_{gb}\\
      &+ 
   C_1 (C_3 + C_a) (C_{ga} + C_{gb}))\large{/}\\
&((C_3 C_4 C_a + 
      C_2 C_3 (C_4 + C_a) + (C_2 + C_3) C_4 C_b\\
      &+ (C_2 + C_3 + C_4) C_a C_b + 
      C_1 (C_3 + C_a) (C_4 + C_b) \\
      &+ 
      C_1 C_2 (C_3 + C_4 + C_a + C_b)) C_{ga}\\
      &+ (C_3 C_4 C_a + C_3 C_4 C_b + 
      C_3 C_a C_b \\
      &+ C_4 C_a C_b + C_3 C_a C_{ga} + C_3 C_b C_{ga} + C_a C_b C_{ga}\\
      &+ 
      C_2 (C_3 + C_b) (C_4 + C_a + C_{ga}) \\
      &+ C_1 (C_3 + C_a) (C_4 + C_b + C_{ga}) \\
      &+ 
      C_1 C_2 (C_3 + C_4 + C_a + C_b + C_{ga})) C_{gb})
\end{aligned}
\end{equation}
The capacitance for qubit A, $C_A$, is as follows:
\begin{equation}
    \begin{aligned}
        C_A&=(2 (C_3 C_4 C_a + 
      C_2 C_3 (C_4 + C_a) + (C_2 + C_3) C_4 C_b \\
      &+ (C_2 + C_3 + C_4) C_a C_b + 
      C_1 (C_3 + C_a) (C_4 + C_b) \\
      &+ C_1 C_2 (C_3 + C_4 + C_a + C_b)) C_{ga} \\
      &+ 
   2 (C_3 C_4 C_a + C_3 C_4 C_b + C_3 C_a C_b + C_4 C_a C_b \\
   &+ C_3 C_a C_{ga} + 
      C_3 C_b C_{ga} + C_a C_b C_{ga}\\
      &+ C_2 (C_3 + C_b) (C_4 + C_a + C_{ga}) \\
      &+ 
      C_1 (C_3 + C_a) (C_4 + C_b + C_{ga})\\
      &+ 
      C_1 C_2 (C_3 + C_4 + C_a + C_b + C_{ga})) C_{gb})/\\
      &(C_4 C_b C_{ga} + 
   C_3 (C_4 + C_b) C_{ga}\\
   &+ C_1 (C_2 + C_4 + C_b) C_{ga} + C_b (C_4 + C_{ga}) C_{gb} \\
   &+ 
   C_3 (C_4 + C_b + C_{ga}) C_{gb}\\
   &+ C_1 (C_2 + C_4 + C_b + C_{ga}) C_{gb} \\
   &+ 
   C_2 (C_3 + C_b) (C_{ga} + C_{gb}))
    \end{aligned}
\end{equation}
The capacitance for qubit B, $C_B$, is as follows: 
\begin{equation}
    \begin{aligned}
        C_B &= (2 (C_3 C_4 C_a + 
      C_2 C_3 (C_4 + C_a) + (C_2 + C_3) C_4 C_b \\
      &+ (C_2 + C_3 + C_4) C_a C_b + 
      C_1 (C_3 + C_a) (C_4 + C_b) \\
      &+ C_1 C_2 (C_3 + C_4 + C_a + C_b)) C_{ga}\\ 
      &+ 
   2 (C_3 C_4 C_a + C_3 C_4 C_b + C_3 C_a C_b + C_4 C_a C_b \\
   &+ C_3 C_a C_{ga} + 
      C_3 C_b C_{ga} + C_a C_b C_{ga} \\
      &+ C_2 (C_3 + C_b) (C_4 + C_a + C_{ga}) \\
      &+ 
      C_1 (C_3 + C_a) (C_4 + C_b + C_{ga})\\
      &+ 
      C_1 C_2 (C_3 + C_4 + C_a + C_b + C_{ga})) C_{gb})/\\
      &((C_2 + C_3) C_4 C_{ga} + (C_2 + 
      C_3 + C_4) C_a C_{ga} \\
      &+ C_a (C_4 + C_{ga}) C_{gb} + C_2 (C_4 + C_a + C_{ga}) C_{gb} \\
      &+ 
   C_3 (C_4 + C_a + C_{ga}) C_{gb} \\
   &+ C_1 (C_2 + C_3 + C_a) (C_{ga} + C_{gb}))
    \end{aligned}
\end{equation}
The capacitance representing the $LC$-mode is as follows:
\begin{equation}
\begin{aligned}
C_{LC} &= (2 (C_3 C_4 C_a + 
      C_2 C_3 (C_4 + C_a) + (C_2 + C_3) C_4 C_b \\
      &+ (C_2 + C_3 + C_4) C_a C_b + 
      C_1 (C_3 + C_a) (C_4 + C_b) \\
      &+ C_1 C_2 (C_3 + C_4 + C_a + C_b)) C_{ga} \\
      &+ 
   2 (C_3 C_4 C_a + C_3 C_4 C_b + C_3 C_a C_b + C_4 C_a C_b \\
   &+ C_3 C_a C_{ga} + 
      C_3 C_b C_{ga} + C_a C_b C_{ga} \\
      &+ C_2 (C_3 + C_b) (C_4 + C_a + C_{ga}) \\
      &+ 
      C_1 (C_3 + C_a) (C_4 + C_b + C_{ga}) \\
      &+ 
      C_1 C_2 (C_3 + C_4 + C_a + C_b + C_{ga})) C_{gb})/\\
      &(((C_3 + C_4) C_a + (C_3 + 
         C_4 + 4 C_a) C_b\\
         &+ C_1 (C_3 + C_4 + C_a + C_b) \\
         &+ 
      C_2 (C_3 + C_4 + C_a + C_b)) C_{ga} \\
      &+ (C_3 C_a + C_4 C_a + C_3 C_b + C_4 C_b + 
      4 C_a C_b \\
      &+ (C_a + C_b) C_{ga} + C_1 (C_3 + C_4 + C_a + C_b + C_{ga})\\
      &+ 
      C_2 (C_3 + C_4 + C_A + C_b + C_{ga})) C_{gb})
\end{aligned}
\end{equation}

To further simplify, with fair approximation of $C_1 \approx C_2$ and $C_3 \approx C_4$, we get the following expressions for each capacitive coupling constants and capacitance terms. Then, the capacitive coupling constant between qubit A and B, $J_C$, becomes as follows:
\begin{equation}
\begin{aligned}
    J_C = &(4e^2)\left[\left(C_3^2-C_1^2\right)\left(C_{ga}+C_{gb}\right)+C_3C_{ga}C_{gb}\right]/\\
&\left[\left(C_1+C_3\right)\left(2C_aC_b+C_3(C_a+C_b)\right.\right.\\
&\left.+C_1(2C_3+C_a+C_b\right))\left(C_{ga}+C_{gb} \right)\\\
&+\left(C_1^2+C_aC_b+C_3(C_a+C_b\right)\\
&\left.+C_1\left(2C_3+C_a+C_b\right))C_{ga}C_{gb}\right]
\end{aligned}
\end{equation}
The capacitive coupling constant between qubit A and $LC$-Mode, $g_{A}$, becomes as follows:
\begin{equation}
    \begin{aligned}
        g_A = &(4e^2)[((C_1+C_b)C_{ga}C_{gb})]/\\
        & [(C_1+C_3)(2C_aC_b+C_3(C_a+C_b)\\
        & + C_1(2C_3+C_a + C_b))(C_{ga}+C_{gb})\\
        &+(C_1^2+C_a C_b+C_3(C_a+C_b)\\
        &+C_1(2C_3+C_a+C_b))C_{ga}C_{gb}]
    \end{aligned}
\end{equation}
The capacitive coupling constant between qubit B and $LC$-Mode, $g_{B}$, becomes as follows:
\begin{equation}
    \begin{aligned}
        g_B = &-(4e^2) [((C_1 + C_a) C_{ga} C_{gb})]/ \\
        & [(C_1 + C_3) (2 C_a C_b + C_3 (C_a + C_b) \\
        &+ C_1 (2 C_3 + C_a + C_b)) (C_{ga} + C_{gb}) \\
        &+ (C_1^2 + C_a C_b + C_3 (C_a + C_b) \\
        &+ C_1 (2 C_3 + C_a + C_b)) C_{ga} C_{gb}]
    \end{aligned}
\end{equation}
It is noteworthy that the definition of $g_A$ and $g_B$ here differ from the $g_A$ and $g_B$ in the main text by a coefficient connecting the charge operator and the creation(annihilation) operators.

The capacitance for qubit A, $C_A$, becomes as follows:
\begin{equation}
    \begin{aligned}
        C_A &= [(C_{ga} + C_{gb})(C_1 + C_3)(2 C_a C_b \\
        &+ C_3 (C_a + C_b) + C_1 (2 C_3 + C_a + C_b))\\
        &+C_{ga} C_{gb}(C_1^2 + C_a C_b + C_3 (C_a + C_b)\\
        &+ C_1 (2 C_3 + C_a + C_b))]/ \\
        &[(C_1 + C_3) (C_1 + C_3 + 2 C_b) (C_{ga}+C_{gb})\\
        &+ (C_1 + C_3 + C_B) C_{ga} C_{gb}]
    \end{aligned}
\end{equation}
The capacitance for qubit B, $C_B$, becomes as follows: 
\begin{equation}
    \begin{aligned}
        C_B &= [(C_1 + C_3)(2 C_a C_b + C_3 (C_a + C_b) \\
        &+ C_1 (2 C_3 + C_a + C_b))(C_{ga} + C_{gb})\\
        &+ (C_1^2 + C_a C_b + C_3 (C_a + C_b) \\
        &+ C_1 (2 C_3 + C_a + C_b))C_{ga}C_{gb}] / \\
        &[(C_1 + C_3) (C_1 + C_3 + 2 C_a) (C_{ga}+ C_{gb})\\
        &+ (C_1 + C_3 + C_A) C_{ga} C_{gb}]
    \end{aligned}
\end{equation}
The capacitance representing the $LC$-mode becomes as follows:
\begin{equation}
    \begin{aligned}
        C_{LC} &= [(C_1 + C_3)(2 C_a C_b + C_3 (C_a + C_b) \\
        &+ C_1 (2 C_3 + C_a + C_b))(Cga + Cgb)\\
        &+ (C_1^2 + C_a C_b + C_3 (C_a + C_b) \\
        &+ C_1 (2 C_3 + C_a + C_b))C_{ga}C_{gb}] / \\ 
        &[2(2C_aC_b + (C_3 + C_1)(C_a + C_b) \\
        &+ 2C_1C_3)(C_{ga} + C_{gb}) \\
        &+ (C_a + C_b + 2C_1)C_{ga}C_{gb}]
    \end{aligned}
\end{equation}
\section{Decoherence analysis}\label{sec:decoherenceanalysis}
\begin{table}[h]
\centering
\caption{\label{tab:coherencetimes}%
System coherence times}
\begin{ruledtabular}
\begin{tabular}{lcc}
\textrm{Decay time} & \textrm{Qubit A} & \textrm{Qubit B} \\
\colrule
\quad \quad{$T_1$ ($\mu$s)}  & {260} & {160} \\
\quad \quad{$T_2^*$ ($\mu$s)}  & {100} & {110} \\
\quad \quad{$T_2^E$ ($\mu$s)}  & {200} & {150} \\
\end{tabular}
\end{ruledtabular}
\end{table}
The coherence times of our system are provided in Table \ref{tab:coherencetimes}. According to our analysis, the decoherence mechanism aligns well with those reported in previous studies that have investigated the same setup \cite{nguyen2019high, somoroff2023millisecond, mencia2024integer}. These studies have identified dielectric loss and thermal photon-induced dephasing as the dominant decoherence mechanisms. We follow the method of \cite{nguyen2019high} to analyze our system, and our calculations further support the understanding, with the extracted effective loss tangent  $\tan \delta_C  \approx 5 \times 10^{-6}$ and cavity temperature $T \approx 55$ mK, showing good agreement with prior results.
\clearpage

\nocite{*}

\bibliography{apssamp}

\end{document}